\documentclass[a4paper,11pt]{article}
\usepackage{pos}
\usepackage{soul}
\sethlcolor{yellow}
\usepackage{amsmath}
\usepackage[varg]{txfonts}
\usepackage[normalem]{ulem}

\def\apj{ApJ}                 
\def\aap{A\&A}                

\def\ssr{Space Science Reviews}

\title{Potential of the Cherenkov Telescope Array for studying the young nearby supernova remnants}
 \ShortTitle{Young nearby SNRs with CTA}

\author*[a]{Dmitry A. Prokhorov}
\author[b]{Shiu-Hang Lee}
\author[c]{Shigehiro Nagataki}
\author[a]{Jacco Vink}
\author[d]{Donald C. Ellison}
\author[e]{Gilles Ferrand}
\author[f]{Daniel J. Patnaude}
\author[g]{Friedrich K. Roepke}
\author[h]{Ivo R. Seitenzahl}
\author[f]{Patrick O. Slane}

\affiliation[a]{GRAPPA, Anton Pannekoek Institute for Astronomy, University of Amsterdam, Science Park 904, 1098 XH Amsterdam, The Netherlands}
\affiliation[b]{Kyoto University, Department of Astronomy, Oiwake-cho, Kitashirakawa, Sakyo-ku, Kyoto 606-8502, Japan}
\affiliation[c]{Astrophysical Big Bang Laboratory (ABBL), RIKEN Cluster for Pioneering Research, 2-1 Hirosawa, Wako, Saitama 351-0198, Japan}
\affiliation[d]{Physics Department, North Carolina State University, Box 8202, Raleigh, NC 27695, USA}
\affiliation[e]{Department of Physics and Astronomy, room 311 Allen Building, 30A Sifton Road, University of Manitoba, Winnipeg, Manitoba, R3T 2N2, Canada}
\affiliation[f]{Harvard-Smithsonian Center for Astrophysics, Cambridge, MA 02138, USA}
\affiliation[g]{Institut f\"{u}r Theoretische Astrophysik, Zentrum f\"{u}r Astronomie, Universit\"{a}t Heidelberg, Philosophenweg 12, 69120 Heidelberg, Germany} 
\affiliation[h]{School of Science, University of New South Wales Canberra, The Australian Defence Force Academy, Canberra, ACT 2600, Australia}

\onbehalf{for the CTA consortium}

\emailAdd{d.prokhorov@uva.nl}

\abstract{Modern imaging atmospheric Cherenkov telescopes have extensively observed young nearby supernova remnants (SNRs), with ages of about 1000 years or less, in the very-high-energy (VHE) gamma-ray band. These efforts resulted in the detection of VHE emission from three young SNRs -- Cassiopeia A, Tycho, and SN 1006 -- and provided significant evidence for emission from the more distant Kepler's SNR. However, many questions on the production of VHE gamma rays in these remnants remain unanswered. Using detailed physical models for Tycho's SNR based on the CR-hydro-NEI code and physically motivated models for the other young nearby remnants, we simulated observations with the Cherenkov Telescope Array (CTA) of these gamma-ray sources. We highlight properties of these remnants accessible for investigation with future CTA observations and discuss which questions are expected to be answered.}

\ConferenceLogo{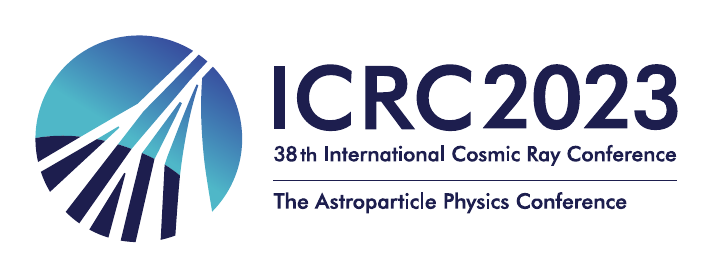}

\FullConference{%
38th International Cosmic Ray Conference (ICRC2023)\\
  26 July - 3 August, 2023\\
  Nagoya, Japan}

\begin{document}
\maketitle

\section{Introduction}

Young ($\lesssim 1000$ yr) supernova remnants (SNRs) provide a record of the most recent stellar explosions \citep[for a review, see][]{Vink2020}.
The rate of supernovae in the Milky Way is approximately one per fifty years. However, due to their predominant location in the dusty plane of the Milky Way, a large fraction of the Galactic supernovae may not have been seen by human eyes. With the exception of SN 1987A, the naked-eye supernovae observed throughout human history, known as historical supernovae, have occurred at relatively close distances of less than 5 kpc.
Such supernovae are luminous optical events, typically powered by the energy input associated with the decay of radioactive nuclei synthesized in the explosion, and their brightness diminishes as the activity of the radionuclides decrease in the weeks and months after the 
explosion. However, as the optical light of the supernovae fades, their debris continue to propagate with high velocities and drive shocks into the interstellar 
medium. After several hundred years, the typical sizes of SNRs reach a few parsecs, which correspond to an angular extension of a few arcminutes at distances for the Cassiopeia A (hereafter, Cas A), Tycho, and Kepler SNRs. 
Even though invisible to the naked eye, the remnants of supernovae still provide us with information on their explosion and evolution.  

The question that attracted a significant amount of attention in the last century is the origin of cosmic rays (CRs), 
see \citep[][]{Ginzburg1964}. Mostly consisting of high-energy protons, CRs reach Earth after diffusive 
propagation in the Galactic halo. The fundamental question is in which sources these particles were accelerated to the observed high energies. 
One of the most natural source candidates are SNRs, since their energy budget is sufficient to account for the CR production if the fraction of SN explosion kinetic energy converted into high-energy particles is about $\sim$10\%. To produce the CR spectrum observed at the Earth, CR sources need to inject a spectrum which is close to (although slightly steeper than) the test particle energy spectrum at strong SNR shocks, $E^{-2}$.
Two additional facts that strongly support that the SNRs are excellent CR source candidates are (1) the presence 
of a characteristic feature, the so-called pion bump, in the $\gamma$-ray spectra indicating that the $\gamma$-ray 
emission from a handful of SNRs is due to hadronic emission \citep[][]{ref_bump}, and (2) the presence of 
accompanying $\gamma$-ray emission from adjacent dense molecular clouds in a few cases, indicating that high-energy 
protons escape from the SNRs \citep[][]{ref_clouds, ref2_clouds}. Collisionless shocks in SNRs are widely accepted as the 
sites of CR acceleration in the energy range below the Galactic CR knee \citep[][]{ref_shocks}.    

Numerical modeling of SNRs allows for the calculation of the evolution from a supernova into a SNR, for example 
\citep[][]{ref_comp}. The information from observations is essential to determine the initial values of simulation 
parameters through a comparison of the observed and modeled physical quantities in the present epoch. The predictive 
power of computer modeling plays a role in understanding what observations are necessary for achieving the scientific 
objectives. The CR-hydro-NEI code models the SNR hydrodynamics modified to include the effects of non-linear diffusive shock acceleration. A full description of the CR-hydro-NEI code can be found in \citep[][]{ref_lee}. 
The modeling of CR production in SNRs, along with the associated coupling (feedback) between CR production and SNR dynamics, is important, but complex. When such a high degree of complexity is involved, toy models can to some extent be useful to complement the computer simulations. In this feasibility study, we use the results of both computer modeling for Tycho's SNR 
\citep[][]{ref_tycho} to demonstrate the concept and toy models for Cas A, Kepler, and SN 1006. 
We use these models in the setup of Monte Carlo (MC) simulations of future Cherenkov Telescope Array (CTA)  
observations. Compared to the currently existing arrays of telescopes operating at very high energies (VHE; >100 GeV), CTA, which will 
consist of a much larger number of telescopes, will achieve unprecedented performance in sensitivity, angular resolution, and energy resolution. The objective of this feasibility study is to assess the prospects of CTA to discover new and currently unknown aspects of the otherwise well-studied nearby young supernova remnants.

\section{Simulated data sets}

To simulate, reduce, and analyze the data we use the software \texttt{ctools} \citep[][]{ref_ctools}, a package developed 
for the analysis of CTA data. Tycho and Cas A are targets in the northern hemisphere for the northern CTA Observatory array (CTAO-North), while Kepler and SN 1006 are in the southern hemisphere in reach of the southern CTA Observatory array (CTAO-South). We use the CTA Instrument Response Functions (IRFs \citep[][]{ref_irfs}; 
prod5-v0.1\footnote{https://zenodo.org/record/5499840}). 
The IRFs were calculated for the planned CTAO-North and CTAO-South that observe an object 
at three zenith angles (20$^{\circ}$, 40$^{\circ}$, 
and 60$^{\circ}$). We used the IRFs optimized for 50 h observation time. 
The prod5-v0.1 version of the IRFs assumes the CTAO arrays in the so-called Alpha configuration, consisting of 
4 large-sized telescopes and 9 medium-sized telescopes for CTAO-North and 14 medium-sized telescopes and 37 small-sized telescopes for CTAO-South. The IRFs used for the simulated data set of each of these SNRs are listed in Table \ref{Tab1}.  
For each SNR, we performed MC simulations to generate 25 different random photon samples
for each astrophysical model and computed the mean value and the standard deviation for measurable quantities. 
In addition to the SNRs, we added 
the cosmic ray background (`CTAIrfBackground') to the source models.

\begin{table}[h]
\begin{center}
\begin{tabular}{| l | l | l | l | l |}
\hline
SNR & Array & Zenith angle & Azimuth  \\ \hline
Tycho & CTAO-North & 40$^{\circ}$ & north \\ \hline
Cas A & CTAO-North & 40$^{\circ}$  & north \\ \hline
Kepler & CTAO-South & 20$^{\circ}$  & north \\ \hline
SN 1006 & CTAO-South & 20$^{\circ}$  & average \\ \hline 
\end{tabular}
	\caption{The IRFs selected for MC simulations with \texttt{ctools}.}
\label{Tab1}
\end{center}
\end{table}

For the MC simulations, we selected observing conditions that can be expected for observations of these SNRs with CTA. Kepler and SN 1006 can be observed from the CTAO-South site at
small zenith angles, while the other two studied remnants, Tycho and Cas A, are at declinations of about +64$^{\circ}$ and 
+58$^{\circ}$ allowing their observations on La Palma (latitude about 28$^{\circ}$ N) at zenith angle larger 
than 36$^{\circ}$ and 30$^{\circ}$. At La Palma, Tycho and Cas~A are visible in the north, and the geomagnetic effect
is a limiting factor for observations at these zenith angles in that direction \citep[][]{ref_geomag}. 
Taking this into account, we checked the performance of the prod5-v0.1 IRFs for this simulation setup by comparing them with 
the IRFs based on alternative quality cuts. We found that the angular resolution for studying Tycho and Cas A is expected to be finer than that from the prod5-v0.1 IRFs. Therefore, the results presented in these proceedings are conservative.

\section{Analyses and results}

In this section, we describe the model and present the feasibility study for each of the four young nearby SNRs. 
We begin with Tycho's SNR, modeled using CR-hydro-NEI computer simulations. After that, we show the results
for the other three selected SNRs. We highlight properties of these SNRs accessible for investigation with CTA.

\subsection{Tycho}

\begin{figure*}
\centering
    \includegraphics[angle=0, width=.7\textwidth]{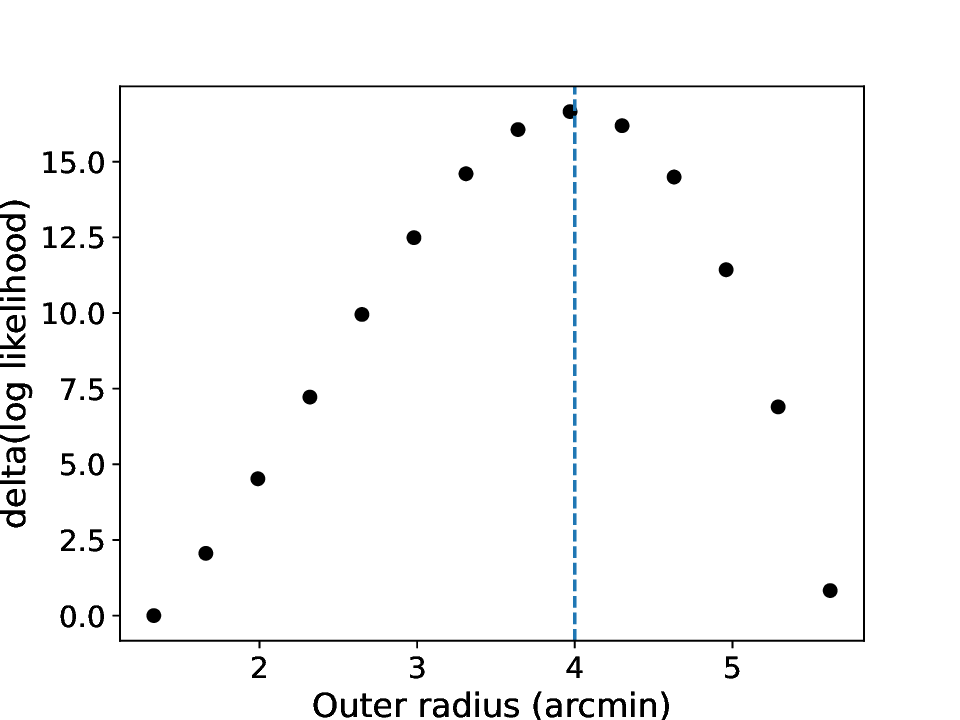}
  \caption{Change in log likelihood while fixing outer radius to values within the range 1.5-5.0 arcmin, as derived from one of the 25 simulated data sets.} 
\label{F1}
\end{figure*}  

Tycho's SNR is the remnant of a the thermonuclear supernova SN 1572. It is therefore just over 450 years old. Along with Cas A and Kepler's SNR, Tycho's SNR 
is one of the youngest Galactic SNRs within a 5 kpc distance. However, being closer than Kepler's SNR and older than Cas A, Tycho's SNR has the largest angular size among the three. In the VHE $\gamma$-ray band, Tycho's SNR was detected with VERITAS \citep[][]{tycho1}. 
Subsequent spatial studies based on the centroid's position of $\gamma$ emission did not reveal a significant shift from the geometrical center of this SNR \citep[][]{tycho2}. We show here that the fine angular resolution of CTA is necessary to distinguish a shell-like morphology from a point-like morphology for this SNR. The addition of Tycho's SNR to the list of the TeV shell-type SNRs, such as Vela Jr., and RCW 86 \citep[][]{listSNR1, listSNR2}, is important, since $\gamma$-ray emission from Tycho's SNR in the TeV band is most likely of a hadronic origin, while $\gamma$-ray emission from the other TeV shell-type SNRs are mostly of a leptonic origin. 

We perform 25 MC simulations for a feasibility study of Tycho's SNR, for which we used the results obtained from the computer simulations, see Model A \citep[][]{ref_tycho}, as input parameters. In this most successful model, the emission is dominated by neutral pion decay resulting from the collisions between relativistic protons accelerated at the forward shock and gas nuclei, with a significant contribution between 1 GeV and 10 GeV arising from inverse Compton scattering of an ambient photon field. In this model, at the current age of Tycho's SNR, the remnant has swept up $\sim$ 2.5 $M_{\odot}$ of ISM material and has converted 16\% of the kinetic energy of the supernova explosion into relativistic particles. The maximum proton energy is nearly 50 TeV. For the radial density profile, the broadband SED covering the $\gamma$-ray band, and the CR proton and electron spectra, see the reference \citep[][]{ref_tycho}. 
To set up these MC simulations, we included two $\gamma$-ray emission components produced via the neutral pion decay and the inverse Compton mechanism, respectively. We binned the modeled emission from Tycho's SNR in one hundred concentric shells in a two-dimensional observational plane. To characterize the spectrum in each radial shell, we used 20 logarithmic energy bins, including 8 energy bins between 100 GeV and 10 TeV. We added a \texttt{ShellFunction} source with an angular width of 3.7 arcsec in the source model for each of the concentric shell and each of the two emission components.

We used three models for a comparison; a background model, a point-like source model, and a shell source model with a width fixed at 1 arcmin and a free shell radius. The point-like source model and the shell source model have power-law spectral shapes. Comparing the background model with the point-like source model, we found the source at the position of Tycho's SNR is expected to be detected at a significance of 35 $\sigma$ with 50 hours of CTAO-North observations. Comparing the shell source model with the point-like source model, we found that the former is expected
to be favored at a significance of 6.5 $\sigma$ using 50 hours of observations. From the simulated data, we checked how precisely the outer radius of the shell and the photon index are expected to be measured and we found confidence intervals of $3.96\pm0.24$ arcmin and $2.21\pm0.03$, respectively. Additionally, we used a disk source model and fitted it to the simulated data. The log likelihood values obtained for shell and disk source models are close and the inferred disk radius is compatible with the inferred outer radius of the shell within statistical errors.
Figure \ref{F1} shows the log likelihood profile for the shell source model. The accurate measurement of the radius is, therefore, feasible and the significance is expected to be sufficient to measure the radius of the shock front at which proton acceleration occurs.    

\subsection{Cas A}

At VHE $\gamma$-ray energies, Cas A is the brightest among the young SNRs. The high $\gamma$-ray 
brightness allows for a spatial analysis on arcmin scales for a point spread function like the one of CTA. Among the four young SNRs 
in our sample, Cas A is the only example of a core-collapse supernova. The revelation of an exponential cut-off at 
a few TeV in its VHE $\gamma$-ray spectrum measured with MAGIC and VERITAS \citep[][]{ref_casaM, ref_casaV} was unexpected, given that it is widely believed 
that young SNRs accelerate CRs up to significantly higher energies. One of the working hypotheses for explaining
the presence of this cut-off is based on a two-zone emission model for regions associated with the forward 
and reverse shocks \citep[][]{ref_front}. In this model, the contribution from the latter is assumed to be dominant at lower energies, while
the contribution from the region associated with the forward shock is still to be discovered above several TeV. We used a disk template and a radial shell template\footnote{The results only slightly depend on the geometry of the latter template.} to model the components associated with
the reverse and forward shocks, respectively. The selected radii correspond to the measured radii of the forward and
reverse shocks \citep[][]{gotthelf}, respectively. We took the spectrum with an exponential cut-off at 2.3 TeV from \citep[][]{ref_casaV}, attributed this spectrum to the region of the reverse shock, and assumed that the component corresponding to the region of the forward shock has a 6 times lower flux at 1 TeV and a power-law spectrum with a photon index of 2.17.       

We fitted three models to the simulated data. The first model is for a point-like source, the second model is for a disk template corresponding to the region associated with the reverse shock, and the third model includes two spatial templates, a radial shell and a disk, for the regions associated with the forward and reverse shocks, respectively. We analyzed simulations with the tool \texttt{CTLIKE} from
\texttt{ctools}, performing a maximum-likelihood fit to the data, and compared the models. The analysis of our MC simulations shows 
that 50-hour observations with CTA will allow a detection of spatially extended emission from Cas A at a significance of 
4.0 $\sigma$ and a detection of the power-law spectral component from the region associated with the forward shock at 
a significance of 5.1 $\sigma$. The former significance depends on the presence of the component associated
with the forward shock in the simulated data and decreases from 4.0 $\sigma$ to 3.5 $\sigma$ in its absence.

\subsection{Kepler}

Kepler's SNR is located at a larger
distance than Tycho's SNR and therefore it is more difficult to detect and perform a spatial morphology study of this source in $\gamma$ 
rays. Gamma-ray emission from Kepler's SNR was measured on the basis of 150 hours of the H.E.S.S. data at VHE energies \citep[][]{KeplerHESS} and on the basis of 12 years of the \textit{Fermi}-LAT data in the GeV band \citep[][]{ref_Xiang, ref_Acero, KeplerHESS}.  
The broad spectral energy distributions of Kepler's SNR and 
Tycho's SNR are similar considering the different distances to these two SNRs \citep[][]{KeplerHESS}. 
Recently developed physical models, discussed above, suggest that the $\gamma$-ray emission from Tycho's SNR is primarily due to hadronic processes. Taking both this fact and the GeV flux from Kepler's SNR into account, it is more than plausible that the hadronic scenario is favored for 
Kepler's SNR as well. To prove the hadronic origin of $\gamma$-ray emission from Kepler's SNR, a spatial morphology analysis with a finer resolution is required. The northern part of Kepler's SNR has a higher gas density than the others parts
of Kepler's SNR \citep[][]{ref_Williams}. If the observed $\gamma$-ray emission is produced in hadronic interaction, then we expect a higher
$\gamma$-ray intensity in the northern part of Kepler's SNR. Given the angular radius of Kepler's SNR of 1.8 arcmin and assuming a dominant contribution from the northern part of the SNR to the total emission, to model the centroid shift we placed a point-like source with the $\gamma$-ray flux taken from the H.E.S.S. paper \citep[][]{KeplerHESS} at the position at 1.3 arcmin offset from the geometrical center of Kepler's SNR.         

We analyzed 25 MC simulations with the tool \texttt{CTLIKE} from \texttt{ctools} and compared the models with a central source and with a source located at an offset of 1.3 arcmin. The analysis shows that 50 hours of CTA observations will result in
a detection of Kepler's SNR at 17 $\sigma$ and that the centroid offset from the geometrical center is expected to be established at 7.5 $\sigma$. We also performed 25 MC simulations lowering the TeV flux normalization by the H.E.S.S. systematic uncertainty and found that the centroid offset from the geometrical center is expected to be revealed at 5.8 $\sigma$.

\subsection{SN 1006}

SN 1006, with a 0.25 deg angular radius, covers a significantly larger fraction of the sky than the other SNRs from our sample. It is projected at a high Galactic latitude of +14.6 deg and exploded into a lower density medium. VHE $\gamma$ rays from this SNR mostly come from its two (north-eastern and south-western) limbs. The spectrum of SN 1006 as observed with the H.E.S.S. array is compatible with a power-law function for each of the limbs \citep[][]{SN1006HESS}. Along with RCW 86, Vela Jr., RX J1731-347, and probably RX J1713-3946, this SNR produces VHE $\gamma$ emission via the inverse Compton scattering by relativistic electrons. The shapes of $\gamma$-ray spectra of RCW 86, Vela Jr., RX J1731-347, and RX J1713-3946 have a hard spectral index in the GeV band and a (super)exponential cut-off at TeV energies. The SN 1006 remnant has a lower $\gamma$-ray luminosity compared with RCW 86, Vela Jr., RX J1731-347, and RX J1713-3946 \citep[e.g.,][]{ref_Condon} and, in contrast, did not
show evidence for a cut-off in the TeV band. To set up MC simulations, we assumed a radial shell morphology with a radius of 0.25 deg and a width of 0.025 deg, and a super-exponential-cut-off power-law spectrum, $\frac{\mathrm{d}N}{\mathrm{d}E}\propto E^{-\alpha} \exp\left(-\sqrt{\frac{E}{/E_{\mathrm{c}}}}\right)$, characteristic for the leptonic scenario from \citep[][]{ref_lept}. Given
the observed bilateral and simulated shell-shaped morphologies, we used a simulated total flux twice as large as the measured one. In this case, 25 hours of simulated data for the whole shell are equivalent to 50 hours of simulated data for the north-eastern and south-western
limbs.

We analyzed 25 MC simulations with \texttt{CTLIKE} from \texttt{ctools} and compared the spectral models with a 
super-exponential-cut-off power-law shape and with a power-law shape. In both analyses the spectral parameters were kept free. Under the assumption of a power-law spectrum, the best-fit index is $2.32\pm0.02$. Under the assumption of a super-exponential-cut-off power-law spectrum, the median best-fit index is $\alpha$=$1.58$ and the median best-fit energy cut-off is $E_{\mathrm{c}}$=$0.69$ TeV, which are close to the model values of $1.56$ and $0.68$ TeV. The uncertainties on the index and the cut-off energy are $0.13$ and $0.22$ TeV, respectively. The spectral curvature in the VHE band 
is expected to be established at a significance of 7.8 $\sigma$.
      
\section{Summary}

We performed a feasibility study of future observations of Tycho's SNR, Cas A, Kepler, and SN 1006 with CTA. The model for Tycho's SNR is based on detailed CR-hydro-NEI computer simulations, while the models for
the other three SNRs are toy models built to incorporate different physical effects. The performed study shows that the spatial extensions of Tycho and Cas A SNRs, the offset of the $\gamma$-ray centroid to the north in 
Kepler's SNR, and the spectral shape of a super-exponential-cut-off in SN 1006 are accessible for investigation with future CTA observations of these young nearby SNRs.  

\section{Acknowledgements}
This work was prepared and reviewed in the CTA Galactic science working group.

\url{https://www.cta-observatory.org/consortium_acknowledgments/}

\end{document}